\documentclass[10pt,aps,pra,twocolumn,floatfix,nofootinbib,superscriptaddress,longbibliography]{revtex4-2}
\usepackage{placeins}

\usepackage{bm,graphicx,mathrsfs,amsmath,amssymb,mathtools,makecell,bbm, amsthm,dsfont,color,times,txfonts,nicefrac,framed,enumitem,tikz,physics,wrapfig,amsfonts,tcolorbox,lipsum,nicematrix}

\usepackage[colorlinks=true,linkcolor=equationcolor,citecolor=teal]{hyperref}

\hypersetup{urlcolor=teal}
\definecolor{equationcolor}{RGB}{222,94,100}
\definecolor{alecolor}{RGB}{198,113,190}
\definecolor{changescolor}{rgb}{0, 0, 0.7}

\makeatletter
\def\blfootnote{\gdef\@thefnmark{}\@footnotetext}
\makeatother

\newcommand{\ms}[1]{\textsf{#1}}
\newcommand{\ot}{\otimes}

\newcommand{\iden}{\mathbbm{1}}

\newcommand{\E}[1]{\mathcal{E}}
\newcommand{\inc}{\mathcal{S}_{\text{inc}}}
\newcommand{\sep}{\mathcal{S}_{\text{sep}}}
\def\A{ {\ms A} }
\def\B{ {\ms B} }

\def\E{ {\cal E} }

\newtheorem{thm}{Theorem}

\newtheorem{lem}[thm]{Lemma}

\begin{document}

\title{Heat as a witness of quantum properties}

\author{A. de Oliveira Junior}
\author{Jonatan Bohr Brask}
	\affiliation{Center for Macroscopic Quantum States bigQ, Department of Physics,
Technical University of Denmark, Fysikvej 307, 2800 Kgs. Lyngby, Denmark}
\date{\today}
\author{Patryk Lipka-Bartosik}
\affiliation{Department of Applied Physics, University of Geneva, 1211 Geneva, Switzerland}
\affiliation{Institute of Theoretical Physics, Faculty of Physics, Astronomy and Applied Computer Science, Jagiellonian University, 30-348 Krakow, Poland}


\begin{abstract}
We present a new approach for witnessing quantum resources, such as entanglement and coherence, based on heat generation. Inspired by Maxwell's demon, we ask what the optimal heat exchange between a quantum system and a thermal environment is when the process is assisted by a quantum memory. We derive fundamental energy constraints in this scenario and show that quantum states can reveal non-classical signatures via heat exchange. This approach leads to a \emph{heat-based} witness for quantum properties, offering an alternative to system-specific measurements, as it only relies on fixed energy measurements in a thermal ancilla. We illustrate our findings with the detection of entanglement in isotropic states and coherence in two-spin systems interacting with a single-mode electromagnetic field.
\end{abstract}

\maketitle


\section{Introduction \label{Sec:introduction}}
Thermodynamics and information theory are fundamentally intertwined. A notable example is Maxwell’s demon~\cite{maxwell2001theory}, an imaginary agent capable of sorting gas molecules by their velocities. By creating a temperature difference, the demon appears to defy the second law of thermodynamics, decreasing the system’s entropy without expending energy. The paradox is resolved by considering the demon’s memory. Erasing the stored information--necessary to complete the cyclic process--inevitably generates heat and increases entropy, as stated by Landauer’s principle~\cite{landauer1961irreversibility,bennett1982thermodynamics,Reeb2014}.

In modern stochastic thermodynamics, Maxwell's demon is regarded as a thermodynamic protocol incorporating measurement and feedback control~\cite{PhysRevLett.100.080403,Esposito_2012,parrondo2015thermodynamics,minagawa2023universalvaliditysecondlaw}. The demon measures the system, stores the outcomes in a memory, and uses them to implement a feedback process. Importantly, the demon's memory is typically considered a classical system, described by orthogonal states.

Experimental realizations of Maxwell's demon on various platforms~\cite{pekola2014,koski2014experimental,Koski2014b,Koski2015, cottet2017observing}, along with advances in both theory~\cite{Rosset_2018} and technology~\cite{lvovsky2009optical,Simon_2010,Heshami_2016} for storing quantum information, raise the question of whether the quantum nature of the memory system can be harnessed in the operation of Maxwell's demon. Beyond storing classical information, ancillary quantum systems can store ``unspeakable" information that cannot be transmitted classically~\cite{marvian2016quantify} and can also become entangled with the system. This leads to a dynamics that cannot be realised with a classical memory~\cite{Rio2011}.

In this work, we consider a quantum version of the Maxwell's demon \cite{cottet2017observing,naghiloo2018information,masuyama2018information}, namely one with a quantum memory that can start in any state. We show that such demons can generate heat flows exceeding classical limits. More precisely, we derive universal and model-independent bounds on heat exchange in thermodynamic processes such as cooling and heating. These bounds highlight a fundamental link between thermodynamic quantities and quantum features. In particular, we show that the quantum nature of the system manifests itself through heat exchange with the surrounding thermal environment, with distinct thermal signatures tied to quantum properties like entanglement and coherence.

\begin{figure}[t]
    \centering
    \vspace{0.3cm}
    \includegraphics{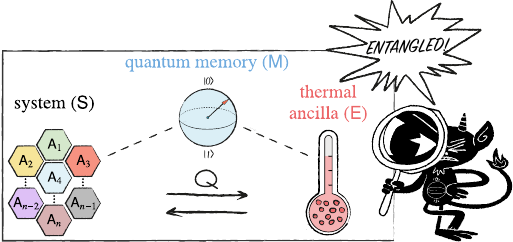}
    \caption{\emph{Heat as a witness}. A device composed of a quantum memory $\ms{M}$ and a thermal ancilla $\ms{E}$ is used to test properties of the state $\ms{S}$. Here, the device determines whether an $n$-partite system $\ms{S} = \A_1...\A_n$ is entangled by measuring the heat flowing into the thermal ancilla.}
    \label{F-scheme}
\end{figure}

Our results presents a novel approach for detecting quantum features via heat exchange in a thermodynamic process (see Fig.~\ref{F-scheme} for a schematic representation). These findings replace system-specific measurements with a simpler energy measurement on a thermal ancilla. To illustrate our results, we discuss two examples. First, we show entanglement detection for two-qubit Werner states. Second, we propose an experimentally feasible protocol for certifying the coherence of spins interacting with a single-mode electromagnetic field. Crucially, our approach differs from recent works on extending Maxwell's demon to the quantum domain~\cite{zurek1984maxwell,Lloyd1997,naghiloo2018information,masuyama2018information} or linking quantum properties with thermodynamics~\cite{Oppenheim2002,brukner2004macroscopic,PhysRevA.70.062113,wiesniak2008heat,singh2013experimental,Khandelwal_2020,Bertulio2021,Pusuluk2021,Imai2023,deOliveiraJunior2024,Pusuluk2024}, which are often valid only for specific models.

\section{Preliminaries \label{Sec:preliminaries}}
We consider a main system $\ms{S}$, a memory system $\ms M$ and an environment $\ms E$~(see Fig.~\ref{F-scheme}). Each subsystem is described by its local Hamiltonian $H_{\ms{X}}$, with dimension $d_{\ms X}$, and is prepared in a state $\rho_{\ms X}$ with $\ms{X} \in \{\ms S, \ms M, \ms E\}$. While the main system and memory start in arbitrary states, the enviroment is initially in a thermal state, $\rho_{\ms E} = \gamma_{\ms{E}}(\beta)$, where $\gamma_{\ms{X}}(\beta) := e^{-\beta H_{\ms{X}}}/Z_{\ms{X}}(\beta)$ with the partition function $Z_{\ms{X}}(\beta) := \tr(e^{-\beta H_{\ms{X}}})$, and $\beta := 1/k_{\text{B}}T$ being the inverse temperature and $k_B$ the Boltzmann constant.

Our aim is to characterize the maximum and minimum heat that the system can exchange with the environment, assisted by a quantum memory. To ensure that heat is exchanged only between $\ms S$ and $\ms E$, we make two assumptions. First, the joint system $\ms{SME}$ is closed and evolves via an energy-preserving unitary process $U$ as~\cite{Janzing2000}
\begin{equation} \label{eq:unitary_evol}
    \eta_{\ms{SME}} = U(\rho_{\ms{S}}\ot \rho_{\ms{M}}\ot \rho_{\ms{E}})U^{\dagger},
\end{equation}
where $U$ satisfies $[U, H_{\ms{S}} + H_{\ms{M}} + H_{\ms{E}}] = 0$. Since there are no extra sources of energy, the energy exchange between $\ms{SM}$ and $\ms{E}$ is regarded as heat $Q := \tr\left[H_{\ms E} (\eta_{\ms E} - \gamma_{\ms E})\right]$, where $\eta_{\ms E} := \tr_{\ms{SM}}[\eta_{\ms{SME}}]$. Second, we demand that the memory $\ms{M}$ does not exchange energy with  $\ms{SE}$. This is achieved by imposing
\begin{equation}\label{Eq:catalytic-constraint}
   \eta_{\ms M}:= \tr_{\ms{SE}}[U(\rho_{\ms{S}}\ot \rho_{\ms{M}}\ot \rho_{\ms{E}})U^{\dagger}] = \rho_{\ms{M}}.
\end{equation}
In other words, the memory affects the system's dynamics but not its energetics. This includes the scenario of Maxwell's demon with feedback~\cite{PhysRevLett.109.180602,ito2015maxwell,seifert2012stochastic}, with the important distinction that here the control system is genuinely quantum. Note that, due to the Naimark's theorem~\cite{nielsen2001quantum}, any measurement and feedback protocol that restores the memory to its original state without requiring work can be implemented by an appropriate choice of $U$ and $\rho_{\ms{M}}$. Consequently, our framework encompasses, but is not restricted to, the standard Maxwell's demon scenario. Finally, due to Eq.~\eqref{Eq:catalytic-constraint}, the memory undegoes a cyclic evolution. While this process can be equivalently viewed as a catalytic evolution~\cite{Datta2022,Patrykreview}, we emphasize that the memory is not reused in the process. For this reason we refer to it as the \emph{memory}.

Importantly, our setup makes no assumptions about the thermalization model, its timescales, or the nature of the environment. Consequently, we allow for any type of (thermal) environment and memory system. Therefore, the following quantity naturally emerges as a measure of the system's fundamental capacity to accept or release heat:
\begin{align}
\label{Eq:optimal-heat}
\begin{split}
Q_{c/h}(\rho_{\ms S}) :=   \underset{H_{\ms E}, H_{\ms M}, U, \rho_{\ms{M}}}{\min/\max} \:\: &\tr[H_{\ms{E}}(\eta_{\ms{E}} - \rho_{\ms{E}})], \\ 
\textrm{s.t.}  \quad\quad &[U, H_{\ms{S}} +H_{\ms{M}} + H_{\ms{E}}] = 0, \\   &\eta_{\ms M}=\rho_{\ms{M}}. 
\end{split}
\end{align}
The index $c/h$ denotes the minimal and maximal heat for cooling and heating the environment, respectively. We will refer to Eq.~\eqref{Eq:optimal-heat} as the \emph{optimal heat exchange}.


\section{Optimal heat exchange with quantum memory \label{Sec:optimal-heat}}
Determining the optimal heat exchange, i.e., solving the problem in Eq.~\eqref{Eq:optimal-heat}, might at first sight appear to be a very complex task. Surprisingly, however, it is possible to obtain a simple, and illustrative solution, namely
\begin{align}
\label{Eq:optimal-heat-exchange}
    Q_{c/h}(\rho_{\ms{S}}) = E[\rho_{\ms{S}}] - E[\gamma_{\ms{S}}(\beta_{c/h})],
\end{align}
where $E[\rho_{\ms{S}}] := \tr[\rho_{\ms{S}}H_{\ms{S}}]$ is the average energy. The inverse temperatures $\beta_{c/h}$ can be interpreted as the system's temperatures \emph{after} the optimal cooling or heating protocol, determined by the roots of the function
\begin{align}\label{Eq:special-function}
    f_{\theta}(x) = \theta - F_{\beta}[\gamma_{\ms{S}}(x)]
\end{align}
for $\theta = F_{\beta}(\rho_{\ms S})$, with the roots $\beta_{c/h}$ chosen such that $\beta_c \leq \beta_h$. The function $F_{\beta}(\rho_{\ms S}) := \tr(\rho_{\ms S} H_{\ms S}) - \beta^{-1} S(\rho_{\ms S})$ is the quantum free energy with $S(\rho) := -\tr(\rho \log \rho)$ being the von Neumann entropy. The proof of this result is technical and can be found in Appendix~\ref{App:optimal-heat-exchange}. Here, we explore its physical implications by examining specific states of the system. 
\begin{figure}[t]
        \centering
	\includegraphics{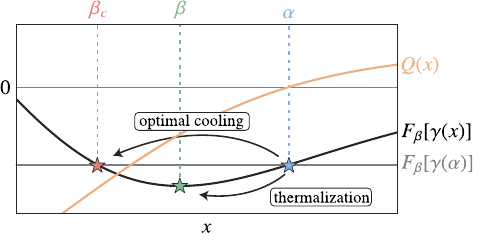}
	\caption{\emph{Cooling using quantum memory}. A quantum system, initially in a thermal state at inverse temperature $\alpha$, interacts with a thermal environment at inverse temperature $\beta$. The orange line depicts the heat exchange $Q(x):= \tr[H_{\ms{S}} \gamma(\alpha)] - \tr[H_{\ms{S}} \gamma(x)]$, between the system and environment as the system's final temperature $x$ varies. The grey/black lines represent the system's initial and final free energy. The protocol that minimizes heat exchange does not simply thermalize the system to the environment's temperature. Instead, it drives the system to a state with the same initial free energy but higher entropy. This higher-energy state enables cooling beyond the limits of a simple thermalization, indicating non-Markovian dynamics.
}
\label{F-smart-thermalisation}
\end{figure}
We begin with the simplest case when the system is in a thermal equilibrium with the environment, $\rho_{\ms S} = \gamma_{\ms{S}}(\beta)$. In this case, the function $f_{\theta}(x) = \log [Z_{\ms{S}}(x) / Z_{\ms{S}}(\beta)]$ has a single root at $x = \beta$. Consequently, $\beta_c = \beta_h = \beta$, and both the minimal and maximal heat exchanges are zero, confirming the intuitive statement that no heat flows between systems in thermal equilibrium.
\begin{figure*}
        \centering
	\includegraphics{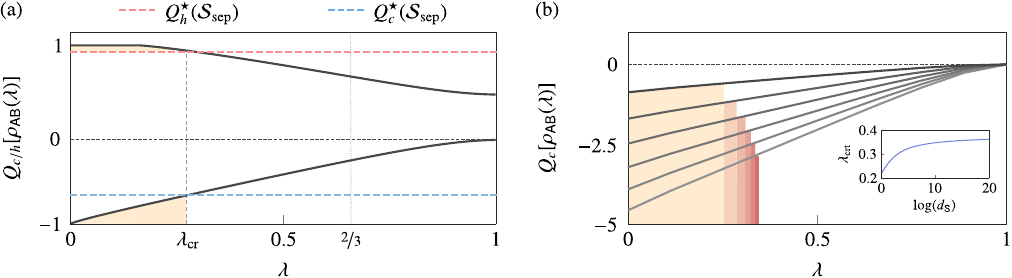}
	\caption{\emph{Entanglement detection via heat exchange}. (a) Heat exchange $Q_{c/h}[\rho_{\ms{AB}}(\lambda)]$ between a two-qubit bipartite system in an isotropic state with an environment at inverse temperature $\beta = 0.5$ (dimensionless), shown as a function of $\lambda$. The state $\rho_{\ms{AB}}(\lambda)$ is entangled for $\lambda < 2/3$ (dashed line). The yellow area represents the range of $\lambda$ where entanglement is detectable by our heat-based witness. The critical parameter $\lambda_{\text{crt}}$ marks the border between negative and non-negative conditional entropy. (b) For general $d$-dimensional isotropic states, detection improves as $\lambda_{\text{crt}}$ increases with $d$. The gradient from black to gray shows the variation in dimension from 2 to 7. The yellow-to-red areas represent entangled states detectable by the heat-based witness $Q_c^{\star}(\mathcal{S}_{\text{sep}})$. The inset shows the behavior of $\lambda_{\text{crt}}$ with the dimension.
}
\label{F:Werner-states-example}
\end{figure*}
If the system is out of equilibrium but remains thermal with respect to a colder temperature $\alpha > \beta$, such that $\rho_{\ms S} = \gamma_{\ms{S}}(\alpha)$, the function $f_{\theta}(x)$ has two roots: $x = \beta_c$ and $x = \beta_h = \alpha$ (see Fig.~\ref{F-smart-thermalisation}). As expected from classical thermodynamics, the optimal heating is zero, $Q_{h}(\rho_{\ms S}) = 0$, meaning the system cannot increase the energy of the environment. This corroborates Clausius' statement of the second law: \emph{heat can never flow from a colder body to a warmer one without some other change} \cite{clausius1865ueber}~(see the orange curve in Fig.~\ref{F-smart-thermalisation} for $\beta = \alpha$). Interestingly, the optimal cooling strategy does not necessarily bring the system to thermal equilibrium. In some cases, as shown by the red star in Fig.~\ref{F-smart-thermalisation}, it occurs when the system's final temperature is hotter than the environment's, with $\beta_c < \beta$. In fact, with no constraints on the process beyond energy conservation, and by allowing a memory to assist, non-Markovian effects emerge, allowing cooling beyond simple thermalization. Similar effects can be observed for the case of heating up the environment, namely when $\alpha < \beta$.

Finally, for a generic quantum state, the effective temperatures $\beta_{c/h}$ depend in a complex way on the density matrix $\rho_{\ms{S}}$. In general, $Q_c(\rho_{\ms S}) \neq Q_h(\rho_{\ms S}) \neq 0$, indicating that a non-equilibrium system can both heat and cool a thermal environment. Thus, $Q_c$ and $Q_h$ provide a quantitative measure of the system’s non-equilibrium nature. This insight is important for our subsequent discussion, as we will show that these two quantities can reveal non-equilibrium properties of quantum states.


\section{Certifying properties of quantum states via heat exchange \label{Sec:certifying}} 
In classical thermodynamics heat exchange is entirely specified by system's macroscopic properties. However, this is not the case in the quantum regime. With access to a quantum memory and the ability to perform arbitrary control, heat exchange becomes sensitive to the fine-grained properties of the system, as demonstrated by Eq.~\eqref{Eq:optimal-heat-exchange}. This leads to the question of whether microscopic properties can be inferred by measuring the heat exchanged with the environment. 

To answer this question, we formalize the problem as follows. Consider a set of density operators $\mathcal{S}$, which we aim to characterize, and define bounds on heat exchange:
\begin{align}\label{Eq:classical-bounds}
 Q_{c}(\mathcal{S}) := \min_{\rho \in \mathcal{S}} Q_{c}(\rho) , \quad \quad  Q_{h}(\mathcal{S}) := \max_{\rho \in \mathcal{S}} Q_{h}(\rho).   
\end{align}
Any state $\rho \in \mathcal{S}$ satisfies $Q_c(\mathcal{S}) \leq Q(\rho) \leq Q_h(\mathcal{S})$. If the system exchanges heat beyond either $Q_{c/h}(\mathcal{S})$, we can conclude it does not belong to the set. Thus, we refer to $Q_{c/h}(\mathcal{S})$ as a \emph{heat-based} witnesses for the set $\mathcal{S}$.

Observe that the optimal heat exchange $Q_{c/h}(\rho)$ crucially depends on the system free energy, as it determines the system temperature after evolution and, consequently, its final state. Therefore, the problem of finding heat-based witnesses for $\mathcal{S}$ can be translated into identifying the state in $\mathcal{S}$ with the highest free energy~(see Appendix~\ref{App:optimal-heat-exchange} for a detailed proof). Formally, let $F_{\beta}(\mathcal{S}) := \max_{\rho \in \mathcal{S}} F_{\beta} (\rho)$ be the largest free energy over $\mathcal{S}$. Then, heat-based witnesses for $\mathcal{S}$ can be expressed as
\begin{align}
    Q_{c/h}(\mathcal{S}) &= E[\rho_{\ms{S}}] - E[\gamma_{\ms{S}}(\beta_{c/h}^{\star})],
\end{align}
where $\beta_{c/h}^{\star}$ is the largest/smallest zero of $f_{\theta}(x)$ for $\theta = F_{\beta}(\mathcal{S})$. 

Remarkably, constructing a heat-based witness only requires measuring the energy of the thermal ancilla before and after the process, without needing a detailed microscopic understanding of the system. This offers a significant advantage over traditional methods, which often rely on entanglement witnesses targeting specific forms of entanglement or require more detailed state information to detect properties like coherence. In contrast, our method bypasses state reconstruction entirely, relying solely on the energy exchange of the thermal ancilla.


\section{Example 1: Entanglement witnessing \label{Sec:example-1}}
We now show how our results can be used to detect any pure bipartite entangled state, as well as mixed entangled states with negative quantum conditional entropy~\cite{cerf1997negative,Cerf1998,friis2017geometry}. Consider a bipartite system $\ms{S} = \ms{AB}$ and let $\mathcal{S}_{\text{sep}}$ be the set of separable states on $\ms{AB}$ with fixed average energy $E_{\ms{AB}}$. Using the techniques described in Sec.~\ref{Sec:certifying}~(see Appendix~\hyperref[App:example-1]{B-1}), we find the optimal heat exchange for $\mathcal{S}_{\text{sep}}$. This leads to a heat-based witness given by $Q^{\star}_{c/h}(\sep) = E_{\ms{AB}} -E[\gamma_{\ms{AB}}(\beta_{c/h}^{\star})]$,
where $\beta_{c/h}^{\star}$ are the zeros of $f_{\theta}(x)$ with $\theta = E_{\ms{AB}} -{\beta}^{-1} \max \{S(\rho_{\ms{A}}), S(\rho_{\ms{B}})\}$.

To illustrate our witness, we consider a family of mixed two-qudit isotropic states with $d_{\ms{A}} = d_{\ms{B}} = d$~\cite{horodecki1999reduction}:
\begin{equation}\label{Eq:isotropic-state}
    \rho_{\ms{AB}}(\lambda) := (1-\lambda)\ketbra{\psi^{+}}_{\ms{AB}}+\lambda\frac{\mathbbm{1}_{\ms{AB}}}{d^2}, 
\end{equation}
where $\lambda \!\in\! [0,1]$, $\ket{\psi^+}_{\ms{AB}}:= d^{-1/2}\sum_{i=1}^d \ket{i}_{\ms{A}}\ket{i}_{\ms{B}}$ and  $\mathbbm{1}_{\ms{AB}}$ is the identity on $\ms{AB}$. Isotropic states are entangled for \mbox{$\lambda < d/(d+1) := \lambda_{\text{ent}}$}~\cite{Horodecki97}.

We first analyze the case of $d=2$, where Eq.~\eqref{Eq:isotropic-state} reduces to Werner states. Figure~\hyperref[F:Werner-states-example]{\ref{F:Werner-states-example}a}, shows the optimal heat exchange $Q_{c/h}[\rho_{\ms{AB}}(\lambda)]$ as a function of $\lambda$ compared to the bounds for separable states $Q^{\star}_{c/h}(\sep)$. As detailed in Appendix~\hyperref[App:example-1]{B-1}, the optimal heat exchange is proportional to the conditional entropy when it is negative, which occurs only for certain entangled states~\cite{friis2017geometry}. Separable states, in contrast, always have positive conditional entropy, with our heat-based witness providing a tight bound when this value is zero. Therefore, we can detect up to a critical value $\lambda_{\text{crt}} \approx 1/4$, where the quantum conditional entropy of $\rho_{\ms{AB}}(\lambda)$ becomes positive. This critical value is specified by a transcendental equation (see Appendix~\hyperref[App:example-1-lambda]{B-2}).

For higher-dimensional systems ($d>2$), the witness improves its detection (see Fig.~\hyperref[F:Werner-states-example]{\ref{F:Werner-states-example}b}, where the colored region below each curve represents the bounds $Q_{c}^{\star}(\sep)$ for that dimension). As the dimension increases, the threshold $\lambda_{\text{ent}}$ also increases, leading to a larger set of entangled isotropic states. Consequently, $\lambda_{\text{crt}}$ increases with $d$, as shown in the inset of Fig.~\hyperref[F:Werner-states-example]{\ref{F:Werner-states-example}b}, making the witness more effective.

We constructed the heat-based witnesses assuming the local entropies and average energy of the system are known. In Appendix~\hyperref[App:example-1-lambda]{B-2}, we show that our results are general and can be extended to cases without knowledge of local entropies. Although we focused on bipartite systems, our results can also be generalised to detect multipartite entanglement (see Appendix~\hyperref[App:example-1-lambda]{B-2}.

\section{Example 2: Certification of Quantum Coherence \label{Sec:examople-2}}
Our second example discusses the construction of a heat-based witness for quantum coherence. Let $\mathcal{S} = \inc$ be the set of incoherent states, defined as those that are diagonal in the energy basis with a fixed average energy $E_{\ms{S}}$. In Appendix \ref{App:example-2}, we show that the optimal heat exchange achievable with states within the set $\inc$ is given by $Q_{c/h}(\inc)  = E_{\ms{S}} - E[\gamma_{\ms{S}}(\beta_{c/h})]\}$, where $\beta_{c/h}$ are the zeros of $f_{\theta}(x)$ with $\theta = E_{\ms{S}} - \beta^{-1} h(E_{\ms{S}}/\norm{H_{\ms{S}}})$, $h(x) := -x\log x -(1-x)\log(1-x)$ is the binary entropy and $\|H_{\ms S}\| = \sup_{\|H_{\ms S}\| = 1} \|H_{\ms S}\|$ is the operator norm. The quantity $Q_{c/h}(\inc)$ can therefore be seen as a heat-based coherence witness for quantum systems of arbitrary dimension.

Let us illustrate how the witness performs in a realistic scenario. Consider two spins, $\ms{S}$ and $\ms{E}$, with equal energy gap $\epsilon$ and energy levels $\{\ket{0}_{\ms{X}}, \ket{1}_{\ms{X}}\}$ for $\ms{X} \in \{\ms{S},\ms{E}\}$, coupled to a single mode electromagnetic field $\ms{M}$ with frequency $\epsilon$. Furthermore, let $a_{\ms{M}}$ denote the bosonic annihilation operator of the field, and $\sigma_{\ms{X}} = \dyad{0}{1}_{\ms{X}}$ be the spin lowering operators. The two spins $\ms{S}$ and $\ms{E}$ should be understood, respectively, as the main system and environment, while the field $\ms{M}$ plays the role of the memory system. The joint system evolves via the Tavis-Cummings Hamiltonian \cite{Tavis1968}, which in the
rotating wave approximation reads
\begin{align}
    H_{\ms{SEM}} = \varepsilon\left(\sigma_{\ms{S}}^{\dagger} \sigma_{\ms{S}} + \sigma_{\ms{E}}^{\dagger} \sigma_{\ms{E}} + a_{\ms{M}}^\dagger a_{\ms{M}}  \right) + V_{\text{int}}, \label{eq:model}
\end{align}
where $V_{\text{int}} :=  g \left(a_{\ms{M}} \sigma_{\ms{S}}^{\dagger} + a_{\ms{M}} \sigma_{\ms{E}}^{\dagger} + \text{c.c.} \right)$, $g > 0$ is the coupling strength and we set $\hbar=1$. Notice that energy is conserved in this setup, and can only be transferred through the memory. 

The system starts in a state with coherence $\rho_{\ms{S}} = \dyad{\psi}_{\ms{S}}$, where $\ket{\psi}_{\ms{S}} \propto \left(\ket{0}_{\ms{S}} + \exp({-\beta H_{\ms{S}}/2})\ket{1}_{\ms{S}}\right)$, and the environment starts in a thermal state $\rho_{\ms{E}} \propto \exp({-\beta \sigma_{\ms{E}}^{\dagger}\sigma_{\ms{E}}})$. The state of the memory is determined as follows. At time $t = 0$ the interaction $V_{\text{int}}$ is turned on and all systems interact for some fixed time $\tau$. This evolution is described by a unitary $U(t) = \exp({-i H_{\ms{SEM}}t})$ and leads to $\eta_{\ms{SEM}}{(t)} = U(t)\left(\rho_{\ms{S}} \ot \rho_{\ms{M}}\ot \rho_{\ms{E}} \right) U^{\dagger}(t)$. Since the field acts as a quantum memory, it must return to its initial state at time $t = \tau$. That is, it must satisfy the operator equation~\eqref{Eq:catalytic-constraint}. This equation fully specifies the initial state $\rho_{\ms{M}}$, which can be determined either analytically or with semi-definite programming~\cite{Boyd2004}.

\begin{figure}[htbp]
    \centering
    \includegraphics{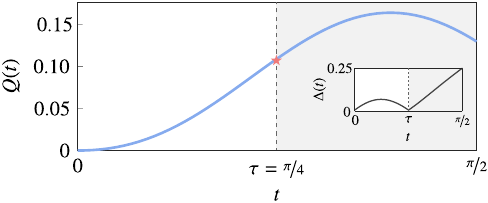}
    \caption{\emph{Coherence detection via heat exchange}. Heat exchange between two atoms is assisted by a single-mode optical cavity, which acts as a quantum memory. The composite system is resonant with $\varepsilon = g=1$. One of the atoms is prepared in a coherent state, whereas the other is in a thermal state at inverse temperature $\beta = 0.3$. The inset shows the change in the cavity state as a function of the distance $\Delta(t):= \|\rho_{\ms{M}}-\eta_{\ms{M}}(t) \|_1$ with respect to the initial state. The red star indicates the final time $\tau = \pi/4$ for which the memory returns to its initial state [$\Delta(\tau) = 0$]. The parameters are chosen such that when $Q(t) > 0$, it corresponds to one of the atoms displaying coherence, whereas $Q(t) = 0$ indicates the absence of coherence.}
    \label{F:coherence-detection}
\end{figure}

The heat exchanged in the process at time $t$ is given by $Q(t) = \varepsilon\text{tr}[\sigma_{\ms{E}}^{\dagger}\sigma_{\ms{E}}\left(\eta_{\ms{E}}(t)-\rho_{\ms{E}}\right)]$, as shown in Fig.~\ref{F:coherence-detection}. When the system is prepared in an incoherent state with energy $E_{\ms{S}}$, no heat can be exchanged, i.e., $Q_{c/h}(\inc) = 0$. This is because the set of incoherent states is given by $\inc = {\gamma_{\ms{S}}(\beta)}$, as the only incoherent state with average energy $E_{\ms{S}}$ is the Gibbs state $\gamma_{\ms{S}}(\beta)$. Therefore, observing $Q(t) \neq 0$ unambiguously indicates the presence of coherence in the initial state of the system. Importantly, during the protocol, the memory state returns to its initial configuration (see inset in Fig.~\ref{F:coherence-detection}). Although the shaded gray area represents the time interval after $\tau$, which is excluded from the analysis, we observe that if the memory state is restored only approximately to its initial state, within some allowable error, the heat exchange remains above our bound, allowing us to witness the presence of coherence.

This example shows a non-classical aspect of the dynamics involving quantum memories where heat exchange would not be possible without energy coherences in the memory system (see Appendix~\hyperref[App:example-1-lambda]{B-2}). This leads to a truly non-classical effect where the quantum nature of the memory provides a way to access the correlations ``locked'' in the non-degenerate energy subspaces, thereby boosting the energy flow between the system and the environment.~\cite{PhysRevLett.130.040401,lipkabartosik2023fundamental}.

\section{Discussion and Outlook \label{Sec:discussion-output}}
In this work, inspired by the thought experiment of Maxwell's demon, we replaced classical memory with a quantum one. This allowed us to establish a fundamental link between thermodynamic quantities and quantum features. In particular, we demonstrated that quantum systems reveal their unique properties through heat exchange with a thermal environment. Our results do not rely on a specific model or particular properties of the system. Instead, they allow us to probe the heat exchange in a quantum process by simply measuring a thermal ancilla. By demonstrating that quantum properties enhance heat exchange, our findings present a viable method for detecting quantum features without the need for full tomographic processes. Although our results were illustrated for detecting entanglement and coherence, they are not limited to these specific properties and can be applied more broadly—from certifying the presence of classical correlations to signaling non-Markovian dynamics.

Our results could be implemented in state-of-the-art experimental setups. For instance, nuclear magnetic resonance (NMR)~\cite{vandersypen2004nmr} and cavity-QED with superconducting qubits~\cite{Blais2021} have been used to implement energy-preserving unitary processes~\cite{Serra2014, serra2016,cottet2017observing}. Additionally both techniques can implement a quantum memory. While NMR enables precise engineering of the system's initial and final states~\cite{Micadei2019}, microcavity systems have been used to mimic quantum memory~\cite{cottet2017observing}, though they have not yet been explored for assisting thermodynamic processes. Other platforms, such as single-electron devices~\cite{Koski2015} or trapped ions~\cite{RevModPhys.75.281,Goold_2014}, are also promising for verifying our results, as they have previously been used to test related ideas. Beyond proof-of-principle experiments, it would be interesting to explore whether the techniques presented here offer practical benefits for detecting quantum features using these platforms.
 
Finally, we highlight that the concepts discussed here can be framed as a catalytic transformation~\cite{Datta2022,Patrykreview}, with the quantum memory being the catalyst or a reference frame~\cite{bartlett2007reference,lostaglio2015description,giacomini2019quantum}. Therefore, our findings demonstrate a concrete application for the concept of quantum catalysis or phase frames~\cite{streltsov2017colloquium}. This contrasts with the typical abstract treatment of catalysts within the resource-theoretic framework~\cite{chitambarGour}.

\begin{acknowledgments}
AOJ and JBB acknowledges financial support from the Danish National Research Foundation grant bigQ (DNRF 142), VILLUM FONDEN through a research grant (40864). P.L.-B. acknowledges the Swiss National Science Foundation (SNSF) through project 192244 and the Polish National Science Centre through project Sonata 2023/37351/D/ST2/02309.
\end{acknowledgments}

\bibliography{2-citations}

\clearpage 
\appendix
\section{Optimal heat exchange \label{App:optimal-heat-exchange}}

We present here the solution to the optimal heat exchange problem as discussed in Sec.~\ref{Sec:optimal-heat}. This begins by demonstrating that the optimization problem posed by Eq.~\eqref{Eq:optimal-heat} can be translated into an equivalent, yet simpler, optimization problem. We then propose an ansatz for the latter and demonstrate that the optimal heat exchange is given by Eq.~\eqref{Eq:optimal-heat-exchange}. The focus then shifts to identifying the state of the system after the process, that is, determining $\gamma_{\ms S}(\gamma_{c/h})$. For clarity, the proof is structured into three distinct steps, outlined below.

\subsection*{Step 1: Equivalence between optimization problems}

We now present a general proof valid for an $n$-partite system, prepared in a state $\rho_{\ms S}$, where $\ms S = \ms{A}_1...\ms{A}_n$, and described by a Hamiltonian $H_{\ms S}:=  \sum_{k=1}^n \mathbbm{1}^{\otimes (i-1)}\otimes H_{\ms A_k}\otimes \mathbbm{1}^{\otimes (n-k)}$. The optimal amount of heat by which the environment can be either cooled down or warmed up is defined by the following optimization problem:
\begin{align}
\label{Eq:app-optimal-heat}
\begin{split}
    Q_{c/h}(\rho_{\ms S}) :=   \underset{H_{\ms E}, H_{\ms M}, U, \rho_{\ms{M}}}{\min/\max} \:\: &\tr[H_{\ms{E}}(\eta_{\ms{E}} - \rho_{\ms{E}})], \\ 
\textrm{s.t.}  \quad\quad &[U, H_{\ms{S}} +H_{\ms{M}} + H_{\ms{E}}] = 0, \\   &\eta_{\ms M}=\rho_{\ms{M}}. 
\end{split}
\end{align}

\subsubsection*{General picture illustrating the equivalence}

Since solving Eq.~\eqref{Eq:app-optimal-heat} is highly non-trivial, we now recast it into a tractable problem. To do so, let us recall some important points. First, the memory can be seen as a catalyst and Eq.~\eqref{Eq:catalytic-constraint} as a catalytic constraint. From now on, we will refer to this equation as the catalytic constraint. Second, due to the catalytic constraint, heat is exchanged only between the main system $\ms S$ and the environment $\ms E$. Third, when $\ms S$ is warmed up, $\ms E$ is cooled down, and vice versa. Consequently, the minimum heat transferred from the system to the environment corresponds to the maximum heat received by the environment. Third, one can interpret such a process from the perspective of a state transformation undergone by the system $\rho_{\ms{S}} \mapsto \eta_{\ms S}$, subject to energy conservation and the catalytic constraint. Putting all the pieces together, we can conclude that this is a well-studied class of quantum channels known as thermal operations~\cite{brandao2015second, horodecki2013fundamental, Lostaglio2019}. In other words, this problem can be framed as a catalytic (coherent) thermal operation that maps $\rho_{\ms{S}}$ to $\tilde{\eta}_{\ms{S}}$, where $\tilde{\eta}_{\ms S} \overset{\epsilon \to 0}{\approx} \eta_{\ms S}$, with $\epsilon$ decaying exponentially with the size of the memory. Therefore, Eq.~\eqref{Eq:app-optimal-heat} can be reformulated as a convex optimisation problem of density operators, namely
\begin{equation}
\begin{aligned}
\label{Eq:app-optimal-heat-2}
    Q_{c/h}(\rho_{\ms S}) := \underset{\eta_{\ms S}}{\min\!/\!\max} \:\: &\tr[H_{\ms{S}}(\rho_{\ms S}-\eta_{\ms S})],\\
&\hspace{-0.95cm}\textrm{s.t.} \:\:\:\:\:\:\:\:  F_{\beta}(\rho_{\ms S}) \geq F_{\beta}(\eta_{\ms S}),
\end{aligned}
\end{equation}
where $F_{\beta}(\rho_{\ms S}) := \tr(H_{\ms S}\rho_{\ms S})-\beta^{-1}S(\rho_{\ms S})$ is the quantum free energy.

\subsubsection*{Formal proof}

To demonstrate the equivalence between Eq.~\eqref{Eq:app-optimal-heat} and Eq.~\eqref{Eq:app-optimal-heat-2}, we proceed in two steps. First, we show that Eq.~\eqref{Eq:app-optimal-heat} is sub-optimal to Eq.~\eqref{Eq:app-optimal-heat-2}, meaning that the optimization constraints of Eq.~\eqref{Eq:app-optimal-heat} can always be satisfied by the broader class of catalytic coherent thermal operations defined in Eq.~\eqref{Eq:app-optimal-heat-2}. This follows naturally because every catalytic coherent thermal operation is a superset of the operations constrained by Eq.~\eqref{Eq:app-optimal-heat}. Second, we establish that any solution achievable under Eq.~\eqref{Eq:app-optimal-heat-2} can also achieve the optimal value of Eq.~\eqref{Eq:app-optimal-heat}. This completes the proof of equivalence. Furthermore, to solidify our claims, we provide explicit constructions demonstrating the achievability of Eq.~\eqref{Eq:app-optimal-heat-2}, emphasizing its practical significance.

We start by proving that the constraints in Eq.~\eqref{Eq:app-optimal-heat} imply $F_{\beta}(\rho_{\ms S}) \geq F_{\beta}(\eta_{\ms S})$. This can be seen by first using the unitary invariance property of the von Neumann entropy, 
$S(\rho_{\ms S})+S(\rho_{\ms M})+S(\rho_{\ms E}) = S(\eta_{\ms{SME}})$, followed by the subadditivity property, which allows us to write the inequality
\begin{equation}\label{Eq:entropic-inequality}
    [S(\eta_{\ms S})-S(\rho_{\ms S})]+[S(\eta_{\ms E})-S(\rho_{\ms E})]\geq 0,
\end{equation}
where $\Delta S_{\ms M} = S(\eta_{\ms M})-S(\rho_{\ms M}) = 0$ due to Eq.~\eqref{Eq:catalytic-constraint}. We now note that the entropy difference $\Delta S_{\ms E}:= [S(\eta_{\ms E})-S(\rho_{\ms E})]$ can be expressed as
\begin{equation}\label{Eq:entropy-difference-reservoir}
   \Delta S_{\ms E} = \beta \tr[H_{\ms E}(\eta_{\ms E}-\rho_{\ms E})]-D(\eta_{\ms E}\|\rho_{\ms E}),
\end{equation}
where $D(\eta_{\ms E}\|\gamma_{\ms E}) := \tr[\eta_{\ms E}(\log \eta_{\ms E} - \log \gamma_{\ms E})]$ is the relative entropy. To show Eq.~\eqref{Eq:entropy-difference-reservoir}, we simply need to recognize that $\gamma_{\ms E}$ is a thermal state (i.e., Gibbs state) and apply straightforward manipulations. 

Next, using the fact that the composite system is closed and that the interaction is energy-preserving, we can write $\tr[H_{\ms E}(\eta_{\ms E}-\rho_{\ms E})] = -\tr[H_{\ms S}(\eta_{\ms S}-\rho_{\ms S})] =: -\Delta E_{\ms S}$. Thus, for $\beta > 0$, Eq.~\eqref{Eq:entropic-inequality} takes the following form:
\begin{equation}\label{Eq:entropy-difference-final}
    \Delta S_{\ms S} -\beta \Delta E_{\ms S}  \geq \Delta S_{\ms S} -\beta \Delta E_{\ms S} - D(\eta_{\ms E}\|\rho_{\ms E}) \overset{\eqref{Eq:entropic-inequality}}{\geq} 0,
\end{equation}
where $\Delta S_{\ms S} := S(\eta_{\ms S}) - S(\rho_{\ms S})$, and we use the observation that the relative entropy is non-negative, $D(\eta_{\ms E}\| \rho_{\ms E}) \geq 0$. Finally, using the definition of the quantum free energy in Eq.~\eqref{Eq:entropy-difference-final}, we obtain that $F_{\beta}(\rho_{\ms S}) \geq F_{\beta}(\eta_{\ms S})$, which holds true for any choice of operators $H_{\ms E}, H_{\ms M}, U$, and $\rho_{\ms{M}}$.

The next step of the proof consists of proving the achievability in Eq.~\eqref{Eq:app-optimal-heat-2}. In other words, we now show that, for sufficiently large memories, there always exists a choice of operators $H_{\ms E}, H_{\ms M}, U$, and a state $\rho_{\ms{M}}$ that achieves the optimum $Q_{c/h}(\rho_{\ms S})$. We begin by considering that the memory is given by the Duan state~\cite{Duan2005,Duan2006}:
\begin{align}
    \label{Eq:duan-state}
    \rho_{\ms{M}}^n = \frac{1}{n}\sum_{i=1}^{n} (\rho^{\ot (i-1)} \ot \eta^{\ot n-i})_{\ms{M}_{\ms S}} \ot \dyad{i}_{\ms{M}_M},
\end{align}
and is described by the Hamiltonian $H_{\ms M} = H^{\ot {n-1}}_{\ms{M}_{\ms{S}}} \oplus \mathbbm{1}_{\ms{M}_M}$, with $\ms{M} = \ms{M}_{\ms{S}} \ms{M}_M$ and $H_{\ms{M}_{\ms{S}}} = H_{\ms{S}}$. We define $\eta^{n-i} := \tr_{1:i}[\eta^n]$, where $\tr_{1:i}[\cdot]$ represents the partial trace over the first $i$ particles, and $\eta^{n}$ is an arbitrary $n$-partite density matrix which will be specified shortly. The unitary $U$ is chosen to be
\begin{align}
    U := U_{\ms{SME}} = W_{\ms{SM}} \left(\sum_{i = 1}^n U^{(i)}_{\ms{S}\ms{M}_{\ms S}\ms{E}} \ot \dyad{i}_{\ms{M}_M}\right), 
\end{align}
where $\{U^{(i)}\}$ is a set of unitaries, such that $U^{(i)} = \mathbbm{1}^{\ot n} \ot \mathbbm{1}_{\ms E}$ for $1 \leq i < n$ and $U^{(n)} = V$, which will be further specified. The unitary $W_{\ms{SM}}$ is a cyclic permutation
\begin{align}
    W_{\ms{SM}}&\left[ \ket{i_1}_{\ms S} \ot \left( \ket{i_2} \ot \ldots \ot \ket{i_n}\right)_{\ms{M}_{\ms{S}}} \ot \ket{i}_{\ms{M}_M}\right] = \ket{i_n}_{\ms S} \ot (\ket{i_1} \ot \ldots \nonumber\\ &\hspace{3cm}\ldots\ot \ket{i_{n-1}})_{\ms{M}_{\ms{S}}} \ot \ket{i+1}_{\ms{M}_M},
\end{align}
with $\ket{n+1}_{\ms{M}_M} \equiv \ket{1}_{\ms{M}_M}$. Note that $U$ leads to the global state $\eta_{\ms{SME}} = U (\rho_{\ms S} \ot \rho_{\ms{M}} \ot \rho_{\ms E})U^{\dagger}$ with $\eta_{\ms{SM}} := \tr_{\ms E}(\eta_{\ms{S}\ms{ME}})$ of the form
\begin{align}
    \label{eq:sigma_AR}
   \!\!\!\! \eta_{\ms{S}{\ms M}} = \frac{1}{n} &\tr_{\ms E} \! \bigg[W_{\ms{S}{\ms M}} \bigg(\rho_{\ms S} \ot \eta^{\ot n-1}_{\ms{M}_{\ms S}} \ot \rho_{\ms E} \ot \dyad{1}_{\ms{M}_M} + \ldots \nonumber\\&\ldots+ \tr_{\ms E}[V(\rho^{\ot n}_{\ms{S} \ms{M}_{\ms S}} \ot\rho_{\ms E})V^{\dagger}] \!\ot\! \dyad{n}_{\ms{M}_M}\bigg) W_{\ms{S}\ms{M}}^{\dagger}\bigg]. 
\end{align}
Observe that $U$ is energy-preserving as long as $V$ is energy-preserving on its support. To complete the achievability proof, we employ the following Lemma:
\begin{lem}[Brandao et. al.~\cite{Brandao2013}]
\label{Lemma:brandao}
Let $\rho$ be an arbitrary density matrix and $\eta$ be an arbitrary block-diagonal density matrix. Then, for any $\epsilon > 0$, there exists an $n\in \mathbb{N}$, a system $\ms E$, with a Hamiltonian $H_{\ms E}$ and a state $\rho_{\ms E}$, along with a unitary  $U$ satisfying $[U, H^{\ot n} + H_{\ms E}] = 0$ and implementing a quantum channel $\mathcal{E}(\cdot) := \Tr_{\ms E}\left[U(\cdot \ot\rho_{\ms E})U^{\dagger}\right]$ such that
\begin{align}
    \label{Eq:app-norm_eq}
    \norm{\mathcal{E}(\rho^{\ot n}) - \eta^{\ot n}}_1 \leq \epsilon.
\end{align}
if and only if $F_{\beta}(\rho) \geq F_{\beta}(\eta)$. Moreover, $\epsilon$ can be taken to be $\sim \mathcal{O}(e^{-a n})$ for some $a \in \mathbb{R}$. 
\end{lem}

From the above result, we know that for any $\epsilon > 0$ there is a sufficiently large $n$, a system $\ms E$ with Hamiltonian $H_{\ms E}$ and state $\rho_{\ms E}$, and an energy-preserving unitary $V$ acting on the subspace $\ms S \ms{M}_{\ms S}\ms E$ such that
\begin{align}
    \label{eq:unitary_from_lemma}
    \norm{\tr_{\ms E}\left[V(\rho^{\ot n}_{\ms S\ms{M}_{\ms S}} \ot \rho_{\ms{E}})V^{\dagger}\right] - \eta^{\ot n}_{\ms S\ms{M}_{\ms S}}}_1 \leq \epsilon.
\end{align}
We choose the unitary from Eq.~\eqref{eq:unitary_from_lemma} to be the one used in our protocol, and the subsystem $\ms E$ to represent the environment, whose existence is guaranteed by Lemma~\ref{Lemma:brandao}. This assures that $U$ is energy-preserving and that the catalytic constraint is safisfied. Finally, the density matrix $\eta^n$ is chosen to be
\begin{align}
    \eta^n := \mathcal{E}(\rho^{\ot n}_{\ms S\ms{M}_{\ms S}}) = \tr_{\ms E}\left[V(\rho^{\ot n}_{\ms S\ms{M}_{\ms S}} \ot \rho_{\ms{E}})V^{\dagger}\right].
\end{align}
As a result, we arrive at 
\begin{align}
    \!\!\!\!\!\!\eta_{\ms S\ms{M}} &\!=\! \frac{1}{n} \left( \eta^n_{\ms S\ms{M}_{\ms S}} \ot \dyad{1}_{\ms{M}_M} + \sum_{i=2}^{n} (\rho^{\ot i-1} \ot \eta^{n-i})_{\ms S\ms{M}_{\ms S}} \ot \dyad{i}_{\ms{M}_M}\right).
\end{align}
A straightforward calculation further shows that
\begin{align}
    \tr_{\ms S}[\eta_{\ms S\ms{M}}] = \rho_{\ms{M}}\quad \text{and} \quad  \norm{\eta_{\ms S\ms{M}} - \eta_{\ms S} \ot \rho_{\ms{M}}}_1 \leq \mathcal{O}(\epsilon),
\end{align}
i.e. the catalytic constraint is satisfied and the protocol generates arbitrarily small correlations between the memory $\ms M$ and system $\ms S$.  Moreover, denoting with $\tr_{/i}(\cdot)$ the partial trace over systems $\{1, \ldots, i-1, i+1, \ldots, n\}$, we can verify that $\tr_{\ms M} (\eta_{\ms S\ms{M}}) = \eta_{\ms S}$ reads
\begin{align}
    \eta_{\ms S} = \frac{1}{n} \sum_{i=1}^n \tr_{/i}\mathcal{E}(\rho_{\ms S}^{\ot n}).
\end{align}
By construction $\mathcal{E}(\rho_{\ms S}^{\ot n})$ is close to $\eta_{\ms S}^{\ot n}$, with $\eta_{\ms S}$ being any state satisfying $F_{\beta}(\rho_{\ms S}) \geq F_{\beta}(\eta_{\ms S})$. Futhermore, Eq.~\eqref{Eq:app-norm_eq} implies that $\tr[\mathcal{E}(\rho_{\ms S}^{n}) -\eta_{\ms S}^{\otimes n}]\leq \epsilon$ for any operator satisfying $0\leq M\leq \iden$. As a result, it follows that
\begin{equation}\label{Eq:heat-system-op}
    \mathcal{E}(\rho_{\ms S}^{\ot n}) \geq \eta_{\ms S}^{\ot n} - \epsilon \mathbbm{1}^{\ot n}
\end{equation}
with $\epsilon \propto \mathcal{O}(e^{-an})$. Therefore, by taking $n$ sufficiently large, which amounts to considering a larger memory system in Eq.~\eqref{Eq:duan-state}, the second term in Eq.~\eqref{Eq:heat-system-op} vanishes and we recover the first line of Eq.~\eqref{Eq:app-optimal-heat}.

\subsection*{Step 2: Implicit solution for the optimization problem}

We now demonstrate that the solution to the optimization problem given by Eq.~\eqref{Eq:app-optimal-heat-2} is given by Eq.~\eqref{Eq:optimal-heat-exchange}. Our analysis begins by examining the constraint in Eq~\eqref{Eq:app-optimal-heat-2}:
\begin{align}
F_{\beta}(\rho_{\ms S}) \geq F_{\beta}(\eta_{\ms S}) \Longleftrightarrow  F_{\beta}(\eta_{\ms S}) - F_{\beta}(\rho_{\ms S}) \leq 0.
\end{align}
The free energy difference $\Delta F_{\beta}:=F_{\beta}(\eta_{\ms S}) - F_{\beta}(\rho_{\ms S})$ can be expressed as

\begin{align}\label{Eq:app-entropy-difference}
\Delta F_{\beta} &= \qty[\tr(H_{\ms S}\eta_{\ms S}) -\frac{1}{\beta}S(\eta_{\ms S})]-\qty[\tr(H_\ms{S} \rho_\ms{S})-\frac{1}{\beta}S(\rho_{\ms S})] \nonumber\\&= \tr[H_{\ms S}(\eta_{\ms S} -\rho_{\ms S})] - \frac{1}{\beta}[S(\eta_{\ms S}) - S(\rho_{\ms S})]. 
\end{align}
Thus, the constraint takes the form of
\begin{align}\label{Eq:app-constraint}
\tr[H_{\ms S}(\eta_{\ms S} -\rho_{\ms S})] \leq \frac{1}{\beta}[S(\eta_{\ms S}) - S(\rho_{\ms S})].
\end{align}

Since there are two solutions for the optimal heat exchange, we will divide the discussion into these two cases for pedagogical purposes. 

We begin with the case where the system warms up, while the environment cools down. By convention, the goal function $\mathcal{Q}:= \tr[H_{\ms S}(\eta_{\ms S}-\rho_{\ms S})] > 0$ is positive as the system warmed up. Now, observe that the left-hand side of the constraint given by Eq.~\eqref{Eq:app-constraint} is the goal function that we aim to maximize. According to Jaynes' principle~\cite{jaynes1957informationI}, when $\eta_{\ms S}$ is a thermal state, it maximizes the entropy $S(\eta_{\ms S})$ and hence maximizes the upper bound of the goal function. However, since we also require the maximization of $Q_c(\eta_{\ms S})$, $\eta_{\ms S}$ must be an inverted thermal state, as this state maximizes the energy while having the same entropy as a thermal state. The difference now is that the population is thermally distributed but completely ordered with respect to the energies. That is, for $E_i \geq E_j$, it implies $p_i \geq p_j$ for all $i$ and $j$ in $\{1, ..., d\}$ -- or, we can simply assume that $\beta_c < 0$. Thus, it follows from the Jaynes' principle that $\eta_{\ms S}$ is given by $\gamma_{\ms S}(\beta_c) = e^{\, \beta_c H_{\ms S}}/\tr[e^{\, -\beta_c H_{\ms S}}]$, with $\beta_{c}$ being the inverse temperature to be determined. Therefore, since the final state of the system is given by $\gamma_{\ms S}(\beta_c)$, we proved one of the two solutions for the optimal heat exchage as given by Eq.~\eqref{Eq:optimal-heat-exchange}. Also, observe that this solution achieves equality in the constraint given by Eq.~\eqref{Eq:app-constraint}. Hence, this allows us to express the optimal heat flow as the entropy difference between the final and initial states, namely $Q_{c}(\rho_{\ms S}) = \beta^{-1}\qty{S(\rho_{\ms S})-S[\gamma(\beta_c)]}$.

The optimal heat exchange in the second case, where the system cools down while the environment warms up, can be proved similarly to the previous case. However, in this case, the left-hand side of the constraint given by Eq.~\eqref{Eq:app-constraint}, which is the objective function we wish to minimize, is now negative. Again, we invoke Jaynes' principle and takes a thermal state as solution $\gamma_{\ms S}(\beta_h) = e^{-\beta_h H_{\ms S}}/Z_{\ms S}.$ Since we aim to minimize the energy this time, the final state is a ``proper'' thermal state. Therefore, we proved the other solution to Eq.~\eqref{Eq:optimal-heat-exchange}.

Importantly, unlike in the previous case, if we wish to express the optimal heat exchange as a function of an entropy difference, we must proceed with caution. This is due to the fact that one cannot cool the system below its ground state, or extract more energy than $E_{\ms S}$ from the system. Therefore, the optimal heat exchange in this case can always be expressed as
$Q_h(\rho_{\ms S}) = \text{min}\qty{\beta^{-1}[S(\rho_{\ms S})-S(\gamma_{\ms S}(\beta_h))],\tr(H_{\ms S}\rho_{\ms S})}$.

\subsection*{Step 3: Finding the inverse temperatures \texorpdfstring{$\beta_{c/h}$}{beta}}\label{App:step-3}

We now show that finding the inverse temperatures boils down to find the roots of a simple non-linear equation. As we saw in the previous subsection, the optimizer is necessarily a Gibbs state of some temperature $x  > 0$, that is $\eta_{\ms{S}} = \gamma_{\ms{S}}(x)$. Second, observe that the optimum is achieved precisely when the constraints are saturated, i.e. $F_{\beta}(\rho_{\ms{S}}) = F_{\beta}(\eta_{\ms{S}})$. Combining these two facts leads to the following equivalent expression for the heat
\begin{equation}
\begin{aligned}
Q_{c/h}(\rho_{\ms S}) := \underset{x}{\min\!/\!\max} \:\: &\tr{H_{\ms{S}}[\rho_{\ms S}-\gamma_{\ms S}(x)]},\\
&\hspace{-0.95cm}\textrm{s.t.} \:\:\:\:\:\:\:\:  F_{\beta}(\rho_{\ms S}) = F_{\beta}[\gamma_{\ms S}(x)],
\end{aligned}
\end{equation}
The temperatures $\beta_{c/h}$ can therefore be found by choosing $x$ that satisfies the equality constraint in the above reformulation. More specifically, defining the function $f_{\theta}(x) = \theta - F_{\beta}[\gamma_{\ms S}(x)]$ for $\theta = F_{\beta}(\rho_{\ms S})$, the constraint is satisfied if and only if $f_{\theta}(x) = 0$. This equation has exactly two solutions (i.e. defining $\beta_c$ and $\beta_h$) whenever $\beta E(\rho_{\ms{S}}) < S(\rho_{\ms{S}})$, one solution $(\beta_c = \beta_h)$ when $\beta E(\rho_{\ms{S}}) = S(\rho_{\ms{S}})$, and no solutions (meaning that the constraint is not saturated) when $\beta E(\rho_{\ms{S}}) > S(\rho_{\ms{S}})$. This last case occurs only for $Q_h(\rho_{\ms{S}})$, and the optimal value of $Q_h(\rho_{\ms{S}})$ is simply given by $E(\rho_{\ms{S}})$.

\section{Details of Example 1 (entanglement witnessing)}

\subsection{Technical details}
\label{App:example-1}
To establish a bound $ F_{\beta}^{\star}(\sep)$ for separable states we observe that for bipartite states the free energy decomposes as $ F_{\beta}(\rho_{\ms{AB}}) = F_{\beta}(\rho_{\ms{A}}) + F_{\beta}(\rho_{\ms{B}})+ \frac{1}{\beta} I(\ms{A}:\ms{B})_{\rho}$, where $I(\ms{A}:\ms{B})_{\rho} :=S(\rho_{\ms{A}})-S(\ms{A}|\ms{B})_{\rho}$ is the quantum mutual information and $S(\A|\B)_{{\rho}}:= S(\rho_\ms{S}) - S(\rho_\B)$ is the quantum conditional entropy~\cite{cerf1997negative,Cerf1998,Cerf1999}. Since $S(\ms{A}|\ms{B})_{\rho}$ is positive for separable states and negative for entangled ones~\cite{Vollbrecht2002}, we find that $F_{\beta}^{\star}(\sep)$ is given by
\begin{align} \label{eq:ent_bound}
     \max_{\hspace{0.3cm}\rho \in \sep} F_{\beta}(\rho)
     &\leq\max_{\hspace{0.3cm}\rho \in \sep} \qty[E(\rho_{\ms{A}}) + E(\rho_{\ms{B}}) - \frac{1}{\beta} \max \{S(\rho_{\ms{A}}), S(\rho_{\ms{B}})\}] \nonumber \\
     &=:  F_{\beta}^{\star}(\sep).
\end{align}
We now show how this leads to a bound on the optimal heat exchange from Eq.~\eqref{Eq:optimal-heat-exchange}. To construct a heat-based entanglement witness, we need to find the optimal heat exchange between the system and the environment when using only separable states. If the observed heat exceeds this limit, it implies that the initial state was entangled. To achieve this, we will leverage the characterization of optimal heat exchange discussed in Section \ref{Sec:optimal-heat}.

In other words, the free energy of any separable state is at most equal to $F_{\beta}^{\star}(\sep)$. Hence, any state that has a higher free energy must have negative conditional entropy, and hence be entangled. To formalize this, we use Eq.~\eqref{Eq:optimal-heat-exchange} and construct an entanglement witness for separable states $\rho_{\ms{AB}}$, namely $Q^{\star}_c(\sep) \leq Q(\rho_{\ms{AB}}) \leq Q^{\star}_h(\sep)$, where
\begin{align}\label{Eq:class-c-h_1}
   \beta Q_{c/h}^{\star}(\sep) = \max\{S(\rho_{\ms{A}}),S(\rho_{\ms{B}})\} -S[\gamma_{\ms{AB}}(\beta_{c/h}^{\star})],
\end{align}
where $\beta_{c/h}^{\star}$ are effective temperatures defined in Sec. \ref{Sec:certifying} and here we opt to write the heat exchange in terms of an entropy difference as done in the previous appendix. This entanglement witness can detect any bipartite pure entangled state of arbitrary finite dimension. This is because the quantum conditional entropy of such states is always negative~\cite{cerf1997negative,Cerf1998,Cerf1999}.  A universal, but loose, bound $F_{\beta}^{\star}(\sep)$ can be obtained by using $S(\rho_{\ms{A}/\ms{B}}) \leq \log d_{\ms{A}/\ms{B}}$ in Eq.~\eqref{eq:ent_bound}.

\subsection{Entanglement witnessing for isotropic and beyond \label{App:example-1-lambda}} 

In this Appendix, we specify the range of parameters $\lambda$ for which the entanglement of isotropic states $\rho_{\ms{AB}}(\lambda)$ can be certified using the witness from Sec.~\ref{Sec:example-1}, and we discuss how our results can be extended to detect multipartite entanglement. 

Observe that the detection of isotropic entangled states exclusively depends on the value $\lambda 
 = \lambda_{\text{crt}}$ for which the conditional entropy $S(\ms{A}|\ms{B})_{\rho}$ becomes negative. That is, the threshold value $\lambda_{\text{crt}}$ corresponds to the case when $S(\ms{A}|\ms{B})_{\rho} = 0$. This can be cast into the following transcendental equation:
\begin{align}
\qty(\lambda_{\text{crt}} -d^2 \lambda_{\text{crt}}) \log \qty(\frac{\lambda_{\text{crt}} }{d^2})&+\qty[d^2 (\lambda_{\text{crt}} -1)-\lambda_{\text{crt}} ] \times \nonumber \\ &\hspace{-1cm}\times\log \qty[\left(\frac{1}{d^2}-1\right) \lambda_{\text{crt}} +1]=d^2 \log d.
\end{align}
The value $\lambda_{\text{crt}}$ which solves the above equation provides the boundary between negative and non-negative conditional entropy. The entanglement witness constructed in Sec. \ref{Sec:example-1} can witness Werner states for which $\lambda < \lambda_{\text{crt}}$. 

Regarding the detection of entanglement in multipartite states, we note that the expression derived for the optimal heat exchange is general and independent of the number of partitions. Below, we generalize all quantities necessary to construct a heat-based entanglement witness for the multipartite scenario.

First, the von Neumann entropy of an $n$-partite system can be written in terms of a chain of conditional entropies. The starting point is to assume that the $n$-partite system is bipartite $\A_1 | \A_2 ... \A_n$, such that $S(\A_1 \A_2...\A_n) = S(\A_2...\A_n) + S(\A_1 | \A_2 ... \A_n)$. By recursively applying this process $(n-1)$ times, we arrive at a general expression for the von Neumann entropy of a multipartite system as a chain of conditional entropies $S(\A_1...\A_n) = S(\A_n) + \sum_{k=1}^{n-1}S\qty(A_k \bigg| \prod_{j=k}^{n-1}A_{j+1})$. Second, we recall that the free energy of a system can be decomposed as $F_{\beta}(\rho_{\ms S}) = \sum_{k=1}^n F_{\beta}({\rho_{\A_k}}) +\beta^{-1}I(\A_1 : \A_2... \A_n)$, where the second term is the mutual information of an $n$-partite system, $I(\A_1 ; \A_2... \A_n) = \sum_{k=1}^n S(\rho_{\A_k}) - S(\rho_{\A_1...\A_n})$. Finally, the free energy of the system can be written in terms of the conditional entropy as $F_{\beta}(\rho_{\ms S}) \!=\! \sum_{k=1}^n F_{\beta}({\rho_{\A_k}}) \!+\!\frac{1}{\beta}\qty[\sum_{k=1}^{n-1} S(\rho_{\A_k}) -\sum_{k=1}^{n-1}S\qty(A_k \bigg| \prod_{j=k}^{n-1}A_{j+1})].$ As before, a heat-based witness for separable states can be derived by noticing that the largest free energy in $\sep$ occurs when the chain of conditional entropies is zero. This leads to the construction of a heat-based witness for multipartite systems.

\section{Details of Example 2 (coherence certification) \label{App:example-2}}

Let $\mathcal{S} = \inc$ be the set of states incoherent (i.e. diagonal) in the energy eigenbasis $\{\ket{i}_{\ms{S}}\}$ with a fixed average energy $E_{\ms{S}} = \tr[H_{\ms{S}} \rho_{\ms{S}}]$. Furthermore, define $\tilde{\rho}_{\ms{S}} = \sum_{i=1}^n \bra{i} \rho_{\ms{S}} \ket{i} \dyad{i}_{\ms{S}}$ to be the state of $\ms{S}$ dephased in its energy basis. Our goal is to find an appropriate bound $F_{\beta}^{\star}(\inc)$ and then use it to construct witnesses $Q_{c/h}^{\star}(\inc)$ for quantum coherence.

We start by observing that $F_{\beta}(\rho_{\ms{S}}) = F_{\beta}(\tilde{\rho}_{\ms{S}}) + A_{\beta}(\rho_{\ms{S}})$, where $A_{\beta}(\rho_{\ms{S}}) := \tr[\rho_{\ms S}(\log \rho_{\ms S}-\log \tilde{\rho}_{\ms S})]$ is the relative entropy of coherence \cite{janzing2006quantum,lostaglio2015description,PhysRevLett.116.120404}.  Crucially, $A_{\beta}(\rho_{\ms{S}}) = 0$ for all incoherent states and positive only for coherent ones. In order to find $\max_{\rho \in \inc} F_{\beta}(\tilde{\rho})$, we note that the maximum is achieved by an incoherent state $\rho^{\star}_{\ms{S}}$ that minimises entropy under fixed average energy $\tr[\rho^{\star}_{\ms{S}}H_{\ms{S}}] = E_{\ms{S}}$.  Such a state is simply given by $\rho_{\ms{S}}^{\star} = (1-p) \dyad{0}_{\ms{S}} + p \dyad{d_{\ms{S}}}_{\ms{S}}$ with $p = E_{\ms{S}}/\epsilon_{d_{\ms{S}}}$. This allows us to find the following (tight) bound
\begin{align}\label{eq:f_incoh}
F_{\beta}^{\star}(\mathcal{S}_{\text{inc}}) &= E_{\ms{S}} - \frac{1}{\beta} h(E_{\ms{S}}/\norm{H}),
\end{align}
where $h(x) := -x\log x -(1-x)\log(1-x)$ is the binary entropy and $\norm{\cdot}$ is the operator norm. Using Eq.~\eqref{eq:f_incoh} we can construct a \emph{heat-based} coherence witness, namely $Q^{\star}_c(\inc) \leq Q(\rho_{\ms{S}}) \leq Q^{\star}_h(\inc) $, where
\begin{align}\label{Eq:class-c-h}
   Q^{\star}_{c/h}(\inc)  = \frac{1}{\beta}\{S(\rho_{\ms S})-S[\gamma_{\ms{S}}(\beta_{c/h}^{\star})]\},
\end{align}
where $\beta_{c/h}^{\star}$ are specified by the zeros of $f_{\theta}(x)$ with \mbox{$\theta = F_{\beta}^{\star}(\inc)$}. 

We start by observing that $F_{\beta}(\rho_{\ms{S}}) = F_{\beta}(\tilde{\rho}_{\ms{S}}) + A_{\beta}(\rho_{\ms{S}})$, where $A_{\beta}(\rho_{\ms{S}}) := \tr[\rho_{\ms S}(\log \rho_{\ms S}-\log \tilde{\rho}_{\ms S})]$ is the relative entropy of coherence \cite{janzing2006quantum,lostaglio2015description,PhysRevLett.116.120404}.  Crucially, $A_{\beta}(\rho_{\ms{S}}) = 0$ for all incoherent states and positive only for coherent ones. In order to find $\max_{\rho \in \inc} F_{\beta}(\tilde{\rho})$, we observe that the maximum is achieved by an incoherent state $\rho^{\star}_{\ms{S}}$ that minimizes entropy under fixed average energy $\tr[\rho^{\star}_{\ms{S}}H_{\ms{S}}] = E_{\ms{S}}$.  Such a state is simply given by $\rho_{\ms{S}}^{\star} = (1-p) \dyad{0}_{\ms{S}} + p \dyad{d_{\ms{S}}}_{\ms{S}}$ with $p = E_{\ms{S}}/\norm{H_{\ms{S}}}$ and $\norm{\cdot}$ being the operator norm. 

\end{document}